\documentclass[preprints,article,accept,moreauthors,pdftex]{Definitions/mdpi} 

\firstpage{1} 
\pubvolume{1}
\issuenum{1}
\articlenumber{0}
\pubyear{2021}
\copyrightyear{2020}
\datereceived{} 
\dateaccepted{} 
\datepublished{} 
\hreflink{https://doi.org/} 

\usepackage{subcaption}
\usepackage{algpseudocode}
\usepackage{algorithm}
\usepackage{physics}

\Title{Channel modeling for in-body optical wireless communications}

\TitleCitation{Channel modeling for in-body optical wireless communications}

\Author{Stylianos E. Trevlakis$^{1,}$ * \orcidA{}, Alexandros-Apostolos A. Boulogeorgos$^2$ \orcidB{}, Nestor D. Chatzidiamantis$^1$ \orcidC{} and George K. Karagiannidis$^1$ \orcidD{}}

\AuthorNames{Stylianos E. Trevlakis, Alexandros-Apostolos A. Boulogeorgos, Nestor D. Chatzidiamantis and George K. Karagiannidis}

\AuthorCitation{Trevlakis, S. E.; Boulogeorgos, A.-A. A.; Chatzidiamantis, N. D.; Karagiannidis, G. K.}

\address{$^1$ Department of Electrical and Computer Engineering, Aristotle University of Thessaloniki, 54124 Thessaloniki, Greece, Emails: $\lbrace$trevlakis; nestoras, geokarag$\rbrace$@auth.gr\\ $^2$ Department of Digital Systems, University of Piraeus, 18534 Piraeus, Greece, Email: al.boulogeorgos@ieee.org}

\corres{Correspondence: trevlakis@auth.gr}

\conference{part at the 10th International Conference on Modern Circuits and Systems Technologies (MOCAST)} 

\abstract{Next generation in-to-out-of body biomedical applications have adopted optical wireless communications (OWCs). However, by delving into the published literature, a gap is recognised in modeling the in-to-out-of channel, since most published contributions neglect the particularities of different type of tissues. Towards this direction, in this paper we present a novel pathloss and scattering models for in-to-out-of OWC links. Specifically, we derive extract analytical expressions that accurately describe the absorption of the five main tissues' constituents, namely fat, water, melanin, oxygenated and de-oxygenated blood. Moreover, we formulate a model for the calculation of the absorption coefficient of any generic biological tissue. Next, by incorporating the impact of scattering in the aforementioned model we formulate the complete pathloss model. The developed theoretical framework is verified by means of comparisons between the estimated pathloss and experimental measurements from independent research works. Finally, we illustrate the accuracy of the theoretical framework in estimating the optical properties of any generic tissue based on its constitution. The extracted channel model is capable of boosting the design of optimized communication protocols for a plethora of biomedical applications.}


\begin{document}
\section{Introduction}
The dawn of the sixth generation (6G) wireless communication era comes with the promise of enabling revolutionary multi-scale applications~\cite{Kaloxylos2020}. An indicative example is in-to-out-of-body biomedical applications~\cite{trevlakis2020commag}. Beyond the technical requirements, such as high reliability and energy efficiency, low power consumption and latency, they need to invest in wireless technologies that from the one hand allow compact deployments, while on the other soothe  possible safety concerns. 

The reliability, speed, energy efficiency and latency of optical wireless communications (OWCs) in biomedical applications have been accurately quantified and experimentally verified over the last decade. Although, a great amount of research effort has been devoted towards optimising in-body OWC systems~\cite{Trevlakis2018mocast,Trevlakis2018mdpi,Trevlakis2018spawc,Trevlakis2019owci,Trevlakis2020aoci}, the current state-of-the-art would greatly benefit from an accurate pathloss model capable of incorporating any generic tissue's characteristics. 

Motivated by this, several researchers turned their eyes to OWCs for biomedical applications. Specifically,~\cite{Zhao2005,Yaroslavsky2002,Zee1993} investigated the optical properties of human brain tissue at various ages in the visible spectrum. Also, in~\cite{Bashkatov2006,Ugryumova2004}, the optical characteristics as well as the mineral density of the bone tissue was measured in the range from $800$ to $2000\text{ }\mathrm{nm}$. Moreover, the authors in~\cite{Sandell2011,Pifferi2004,Spinelli2004} performed experiments in order to measure the optical properties of human female breast tissues in multiple wavelengths and over different distances, while, in~\cite{Tseng2011,Salomatina2006,Shimojo2020}, the optical properties of both healthy and cancerous skin were studied in the visible and near-infrared spectral range. From the aforementioned works, it is observed that the majority of published works has focused on quantifying the optical characteristics of specific tissues at certain wavelength~\cite{Tseng2011,Salomatina2006,Shimojo2020,Sandell2011,Pifferi2004,Spinelli2004,Bashkatov2006,Ugryumova2004,Zhao2005,Yaroslavsky2002,Zee1993}. Despite the significance of these results, they cannot always be exploited from future researchers due to the fact that they may not include all the necessary wavelengths, while even if the desired wavelength is available, the constitution of a tissue is different enough between distinct individuals that the results cannot be regarded as confident.
\vspace{0.2cm}

{\centering
	\begin{minipage}[b]{0.3\textwidth}
		\centering
		\includegraphics[width=1\linewidth,trim=0 0 0 0,clip=false]{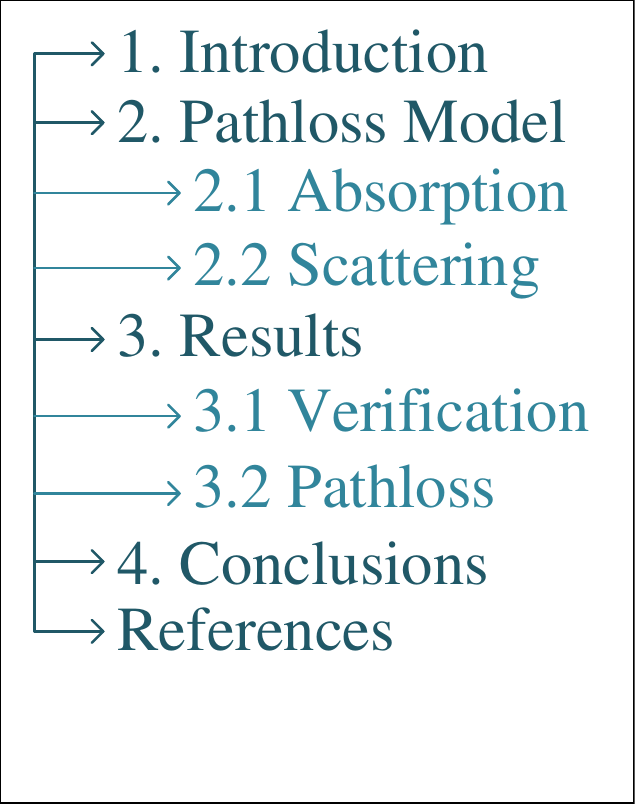}
		\captionof{figure}{The structure of this paper.}
		\label{fig:structure}
	\end{minipage}
	\begin{minipage}[b]{0.4\textwidth}
		\centering
		\begin{tabular}[h]{r|l}
			Variable & Name\\
			\hline
			\hline
			$B$					& Blood volume fraction\\
			$\cos(\cdot)$		& Cosine function\\
			$\delta$			& Tissue thickness\\
			$\exp(\cdot)$		& Exponential function\\
			$F$					& Fat volume fraction\\
			$f_{\text{Ray}}$	& Rayleigh scattering factor\\
			$g$					& Anisotropy factor\\
			$L$					& Pathloss\\
			$M$					& Melanin volume fraction\\
			$\mu_a$				& Absorption coefficient\\
			$\mu_s$				& Scattering coefficient\\
			$\mu'_s$			& Reduced scattering coefficient\\
			$\sin(\cdot)$		& Sine function\\
			$W$					& Water volume fraction\\
		\end{tabular}
		\captionof{table}{Table of variables.}
		\label{fig:variables}
	\end{minipage}
}

\vspace{0.2cm}
A method that estimates the optical properties of any generic tissue based on its constitution is required in order to aid the development of novel biomedical applications that utilise the optical spectrum for communication. In this direction, specific formulas have been reported for the pathloss evaluation of a generic tissue that take into account the variable amounts of its constituents (i.e. blood, water, fat, melanin), but require their optical properties at the exact transmission wavelength, which hinders the use of these formulas~\cite{Jacques2013,TrevlakisOvsE}. The development of such a method will open the road towards not only the theoretical analysis of in-to-out-of body OWC links but also the design of novel transmission and reception schemes, as well as scheduling and routing techniques for next generation networks. Motivated by this, this work derives a novel mathematical model, which requires no experimental measurements for the calculation of the pathloss for in-body OWCs. In more detail, the technical contribution of the this work is summarized as follows:
\begin{itemize}
	\item The absorption coefficients of oxygenated and de-oxygenated blood, water, fat, and melanin are modeled and their analytical expressions are derived.
	\item A general model for calculating the absorption coefficient of any generic tissue is formulated.
	\item We incorporate the impact of scattering expressed in terms of the scattering and reduced scattering coefficients into the aforementioned model.
	\item Numerical results are drawn for both the reduced scattering and the absorption coefficients of complex human tissues, namely breast, skin, bone and brain.
	\item Simulation results from the developed mathematical framework are compared against experimental results from published experimental works. Thus the performance and validity of the theoretical analysis is proven.
	\item Lastly, the pathloss is plotted against the transmission wavelength, the complex tissue types and tissue thickness, while we provide insightful conclusions concerning the future communication protocols.
\end{itemize}

The organisation of this paper as well as a list of variables alongside their names are presented in Fig.~\ref{fig:structure} and Table~\ref{fig:variables}, respectively. In more detail, Section~\ref{S:pathloss_model} is devoted in presenting the pathloss model based on the absorption and scattering properties of the constituents of any generic tissue. Section~\ref{S:results} presents respective numerical results that verify the mathematical framework and insightful discussions, which highlight design guidelines for communication protocols. Finally, closing remarks are summarized in Section~\ref{S:conclusion}. 
 
\section{Pathloss Model} \label{S:pathloss_model}
The losses due the propagation of optical radiation through any biological tissue can be expressed, based on classic OWC theory, as in~\cite{Saidi1995}
\begin{align}
	L = \exp\left(\left(\mu_a + \mu_s\right) \delta\right) ,
\end{align}
where $\delta$ is the propagation distance, while $\mu_s$ and $\mu_a$ represent the scattering and absorption coefficient, respectively. 

\subsection{Absorption} \label{S:absorption}
The absorption coefficient can be modeled as
\begin{align} \label{eq:mu_1}
	\mu_a = -\frac{1}{T} \frac{d T}{d \delta} ,
\end{align}
with $T$ denoting the fraction of residual optical radiation at distance $\delta$ from the origin of the radiation. Thus, the fractional change of the incident light's intensity can be written as in~\cite{coddington1955theory}
\begin{align}
	T = \exp\left(-\mu_a \delta\right) .
\end{align}
As a result, $\mu_a$ can be expressed as the sum of all the tissue's constituents, namely water, melanin, fat, oxygenated blood and de-oxygenated blood. Thus,~\eqref{eq:mu_1} can be analytically expressed as in~\cite{Jacques2013}
\begin{align} \label{eq:mu_a}
	\mu_a = &B S {\mu_a}_{(oBl)} + B \left(1-S\right) {\mu_a}_{(dBl)} + W {\mu_a}_{(w)} + F {\mu_a}_{(f)} + M {\mu_a}_{(m)} ,
\end{align}
where ${\mu_a}_{(i)}$ represents the absorption coefficient of the $i$-th constituent, while $B$, $W$, $F$, and $M$ represent the blood, water, fat, and melanin volume fractions, respectively. Finally, $S$ denotes the oxygen saturation of hemoglobin. 

From~\eqref{eq:mu_a}, we observe that the absorption coefficients and volume fractions of each constituent are required in order to calculate the complete absorption coefficient of any tissue. At this point, it is essential to highlight the dependence of any tissue’s optical properties on the type of tissue, person and even the procedure of collecting the sample, which hinders their mathematical modeling despite of the recent advances in optical measurement techniques. Although such variations are inherent characteristics of each individual and are subject to  tissue preparation protocols, it has been proven in various past publications that they are mainly dependent on the wavelength of the transmitted optical radiation.

\begin{algorithm}[h]
	\caption{ML-based fitting mechanism}
	\begin{algorithmic}[1]
		\Procedure{}{$\{\boldsymbol{{\mu_a}_{(i)}},\mathbf{X}\}_{i\in\{\text{oBl, bBl, w, f}\}},\text{NMSE}, \lambda, {\{a_j, b_j, c_j, w_j\}}_{j\in[0,k_i]}$} 
		\For{$i\in\{\text{oBl, bBl, w, f}\}$}
			\State $k_i \leftarrow \text{degrees of freedom}$
			\For{$j \in[1,k_i]$}
				\State $a_j \leftarrow 0, b_j \leftarrow 0, c_j \leftarrow 1, w_j \leftarrow 0, \lambda \leftarrow 400\text{ nm}.$
				\While{$\{a_j, b_j, c_j, w_j\} \in\mathbb{R}, \lambda \in[400,1000]\text{ nm}$}
					\State $\text{Normalized Mean Square Error (NMSE)}' \leftarrow \frac{\|\boldsymbol{X}-\boldsymbol{{\mu_a}_{(i)}}\|^2_2}{\|\boldsymbol{X}\|^2_2}$
					\If{$\text{NMSE}' < \text{NMSE}$}
						\State \textbf{break}
					\EndIf
				\EndWhile
			\EndFor
		\EndFor
		\EndProcedure
	\end{algorithmic}
	\label{alg:fitting}
\end{algorithm}
Based on the aforementioned, we utilised the non-linear regression-based machine learning algorithm presented in~Algorithm~\ref{alg:fitting}, to derive analytical expressions for the absorption coefficients based on experimental datasets for each of the constituents~\cite{boulogeorgos2020MLinNanoBio}. In more detail, the experimental data of the absorption coefficient of water in~\cite{Hale1973,Zolotarev1969,Segelstein1981} were used to extract a Fourier series that accurately describes the absorption coefficient as a function of the transmission wavelength, which can be written as
\begin{align} \label{eq:m_water}		
	{\mu_a}_{(w)}\left(\lambda\right) = {a_0}_{(w)} + \sum_{i=1}^{7} {a_i}_{(w)} \cos(i w \lambda) + {b_i}_{(w)} \sin(i w \lambda) .
\end{align}
Moreover, the analytical expressions of the absorption coefficients of oxygenated and de-oxygenated blood, were fitted on the experimental results presented in~\cite{Takatani1979,Moaveni1970,Schmitt1986,Zhao2017} by using the sum of Gaussian functions and can be expressed as 
\begin{align} \label{eq:m_oBlood}
	{\mu_a}_{(oBl)}\left(\lambda\right) = \sum_{i=1}^{5} {a_i}_{(oBl)} \exp\left(-\left(\frac{\lambda-{b_i}_{(oBl)}}{{c_i}_{(oBl)}}\right)^2\right) ,
\end{align} 
and 
\begin{align} \label{eq:m_dBlood}
	{\mu_a}_{(dBl)}\left(\lambda\right) = \sum_{i=1}^{4} {a_i}_{(dBl)} \exp\left(-\left(\frac{\lambda-{b_i}_{(dBl)}}{{c_i}_{(dBl)}}\right)^2\right) .
\end{align}
Furthermore, special attention is required for the appropriate preparation of fat tissue, which requires proper purification and dehydration before measuring its optical properties. As mentioned in~\cite{Bashkatov2005}, this necessary procedure can result in inconsistencies between different published works. To limit this phenomenon, we selected the measurements in~\cite{Bashkatov2005} as they coincide with multiple published works in the visible spectrum. The extracted analytical expression of the absorption coefficient of fat can written as a sum of Gaussian functions as follows
\begin{align} \label{eq:m_fat}
	{\mu_a}_{(f)}\left(\lambda\right) = \sum_{i=1}^{5} {a_i}_{(f)} \exp\left(-\left(\frac{\lambda-{b_i}_{(f)}}{{c_i}_{(f)}}\right)^2\right) .
\end{align}
Note that the parameters of the aforementioned expressions are presented in Table~\ref{tbl:parameters}. Finally, the absorption coefficient of melanin is highly consistent throughout various experimental measurements in the visible spectrum~\cite{Jacques1991,Zonios2008,Jacques2013}, based on which we write its analytical expression as
\begin{align} \label{eq:m_melanin}
	{\mu_a}_{(m)}\left(\lambda\right) = {\mu_a}_{(m)}\left(\lambda_0\right) \left(\frac{\lambda}{550}\right)^{-3} ,
\end{align}
with ${\mu_a}_{(m)}\left(\lambda_0\right)$ denoting the absorption coefficient of melanin at $\lambda_0=550\text{ }\mathrm{nm}$ that is equal to $519\text{ }\mathrm{cm}^{-1}$~\cite{Zonios2008}.
\begin{table} [h]
	\centering
	\caption{Fitting parameters for constituent's absorption coefficient.}
	\begin{tabular}{|c|c|c|c|c|}
		\hline
		& dBlood & oBlood & water & fat \\
		\hline
		$a_0$ & - & - & $ 324.1 $ & - \\
		\hline
		$a_1$ & $ 38.63 $ & $ 14 $ & $ 102.2 $ & $ 33.53 $ \\
		\hline
		$a_2$ & $ 60.18 $ & $ 13.75 $ & $ -568 $ & $ 50.09 $ \\
		\hline
		$a_3$ & $ 25.11 $ & $ 29.69 $ & $ -126.6 $ & $ 3.66 $ \\
		\hline
		$a_4$ & $ 2.988 $ & $4.317 \times 10^{15}$ & $ 236.8 $ & $ 2.5 $ \\
		\hline
		$a_5$ & - & $ -34.3 $ & $ 73 $ & $ 19.86 $ \\
		\hline
		$a_6$ & - & - & $ -40.53 $ & - \\
		\hline
		$a_7$ & - & - & $ -12.92 $ & - \\
		\hline
		$b_1$ & $ 423.9 $ & $ 419.7 $ & $ 697.9 $ & $ 411.5 $ \\
		\hline
		$b_2$ & $ 31.57 $ & $ 581.5 $ & $ 121.7 $ & $ 968.7 $ \\
		\hline
		$b_3$ & $ 559.3 $ & $ 559.9 $ & $ -395.3 $ & $ 742.9 $ \\
		\hline
		$b_4$ & $ 664.7 $ & $ -25880 $ & $ -107.1 $ & $ 671.2 $ \\
		\hline
		$b_5$ & - & $ 642.6 $ & $ 115.6 $ & $ 513.8 $ \\
		\hline
		$b_6$ & - & - & $ 35.46 $ & - \\
		\hline
		$b_7$ & - & - & -$ 8.373 $ & - \\
		\hline
		$c_1$ & $ 33.06 $ & $ 16.97 $ & - & $ 38.38 $ \\
		\hline
		$c_2$ & $ 660.8 $ & $ 11.68 $ & - & $ 525.9 $ \\
		\hline
		$c_3$ & $ 59.08 $ & $ 46.71 $ & - & $ 80.22 $ \\
		\hline
		$c_4$ & $ 28.53 $ & $ 4668 $ & - & $ 32.97 $ \\
		\hline
		$c_5$ & - & $ 162.5 $ & - & $ 119.2 $ \\
		\hline
		$w$ & - & - & $ 0.006663 $ & - \\
		\hline
	\end{tabular}
	\label{tbl:parameters}
\end{table}

\subsection{Scattering} \label{S:scattering}
The analysis so far neglects the effects of scattering on the propagation of the optical radiation through the human tissue. So far, the transmission distance, $\delta$, was regarded as a single linear path through the material. However, when taking into account the optical scattering, it becomes the sum of all the paths between scattering events. The analytical evaluation of scattering is not well defined in the literature due to its stochastic nature, especially in diffuse reflectance geometries like the majority of tissues found inside the human body. In more detail, the scattering phenomenon can be decomposed into two factors, namely mean free path and phase function. The former represents the mean distance between two scattering instances, while the latter quantifies the stochastic change in direction of a photon when scattering occurs. Throughout the literature, the mean free path is represented as the inverted scattering coefficient, i.e., $\frac{1}{\mu_s}$, which represents the probability of scattering as a function of the transmission distance. Furthermore, the scattering coefficient can be expressed in terms of the more tractable reduced scattering coefficient as 
\begin{align}
	\mu'_s = \left(1-g\right)\mu_s,
\end{align}
where $g$ represents the anisotropy factor of the generic tissue. On the other hand, the stochastic behavior of the phase function can be mathematically defined based on Mie scattering theory.

Unlike absorption, there is no well-defined analytical relationship that can be used to define the spectral dependence of volume scattering. However, based on experimental observations and simplifying analysis of Mie scattering theory, there is a consensus that the reduced scattering coefficient can be modeled using a power law relationship as in~\cite{Jacques2013}
\begin{align}
	\mu'_s = \mu'_s \left(\delta_0\right) \left(\frac{\delta}{\delta_0}\right)^{-\beta} ,
\end{align}
where $\mu'_s \left(\delta_0\right)$ denotes the reduced scattering coefficient at distance $\delta_0$ and $\beta$ is a dimensionless variable that provides an estimation of the average size of particles in the medium. For instance, if the tissue is composed of small sized particles, $\beta \approx 4$, which can be modeled as Rayleigh scattering. On the contrary, if larger particles exist in the tissue, $\beta$ approaches $0.37$~\cite{Mourant1997,Mourant1998}. However, when a generic tissue is comprised of variable size particles, which is the most common case, the reduced scattering coefficient can be expressed as in~\cite{Bashkatov2005,Lau2009,Saidi1995}
\begin{align} \label{eq:reduced_scattering_coefficient}
	\mu'_s = \mu'_s \left(\delta_0\right) \left( f_{\text{Ray}} \left(\frac{\delta}{\delta_0}\right)^{-4} + \left(1-f_{\text{Ray}}\right) \left(\frac{\delta}{\delta_0}\right)^{-\beta}\right) ,
\end{align}
which corresponds to the combination of the two limiting cases, namely Rayleigh and Mie scattering, with $f_{\text{Ray}}$ denoting the fraction of Rayleigh scattering due to the existence of small sized particles. 

\section{Results} \label{S:results}
This section illustrates the performance of the theoretical framework presented in Section~\ref{S:pathloss_model}. Initially, the extracted expressions for the absorption and scattering coefficients are verified by comparison with experimentally verified results from previously published works. Next, the accuracy of the extracted mathematical framework for estimating the optical properties of any generic tissue is validated against experimental measurements of complex biological tissues, such as skin, bone, breast, and brain tissue. Lastly, the complete pathloss for each of the aforementioned tissues is evaluated and important design guidelines for communication protocols are derived through insightful discussions.
\begin{figure}[h]
	\centering
	\begin{subfigure}[b]{0.36\textwidth}
		\centering
		\includegraphics[width=1\linewidth,trim=0 0 0 0,clip=false]{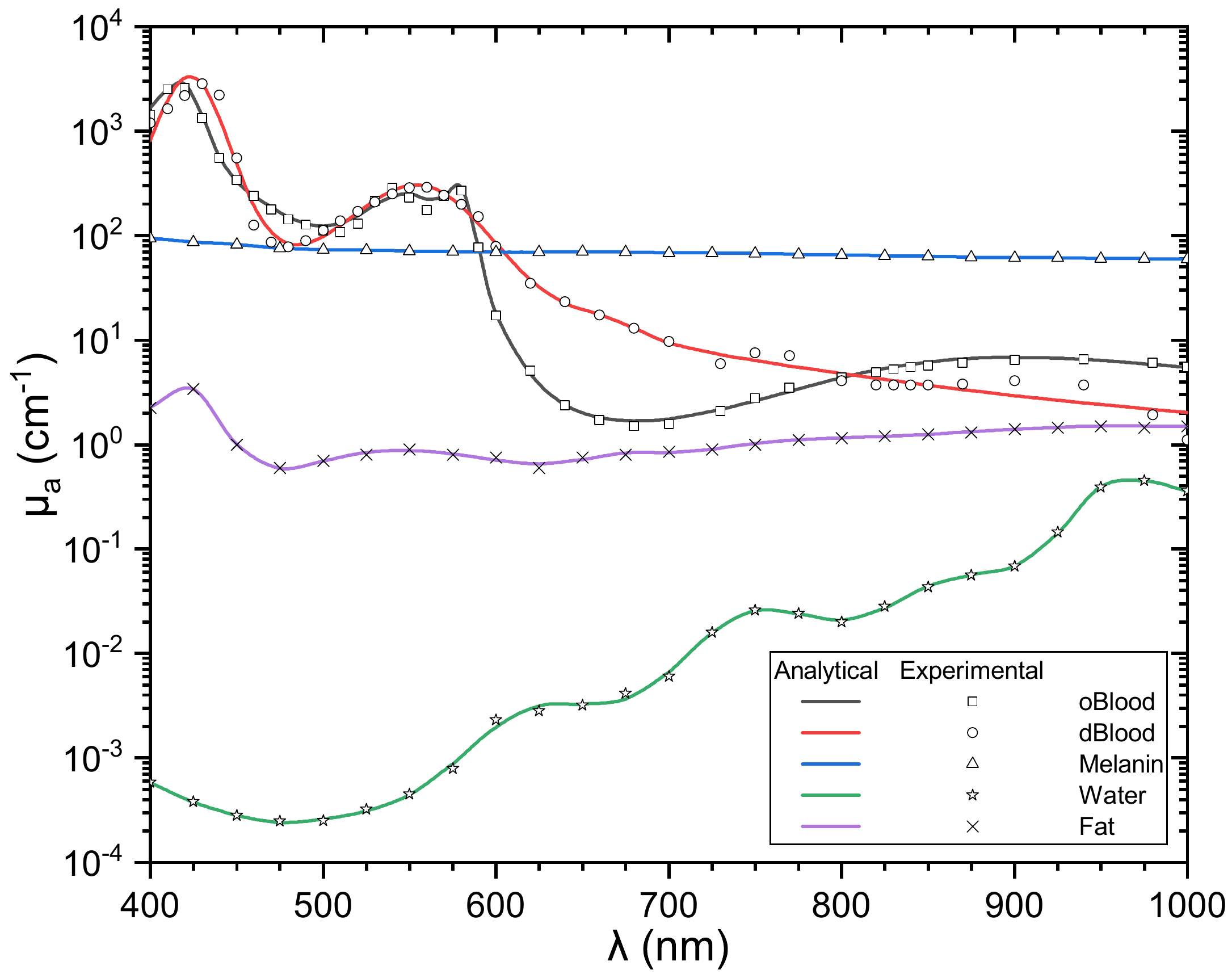}
		\caption{Absorption coefficient of tissue constituents.}
		\label{fig:complete_fitting}
	\end{subfigure}
	\begin{subfigure}[b]{0.36\textwidth}
		\centering
		\includegraphics[width=1\linewidth,trim=0 0 0 0,clip=false]{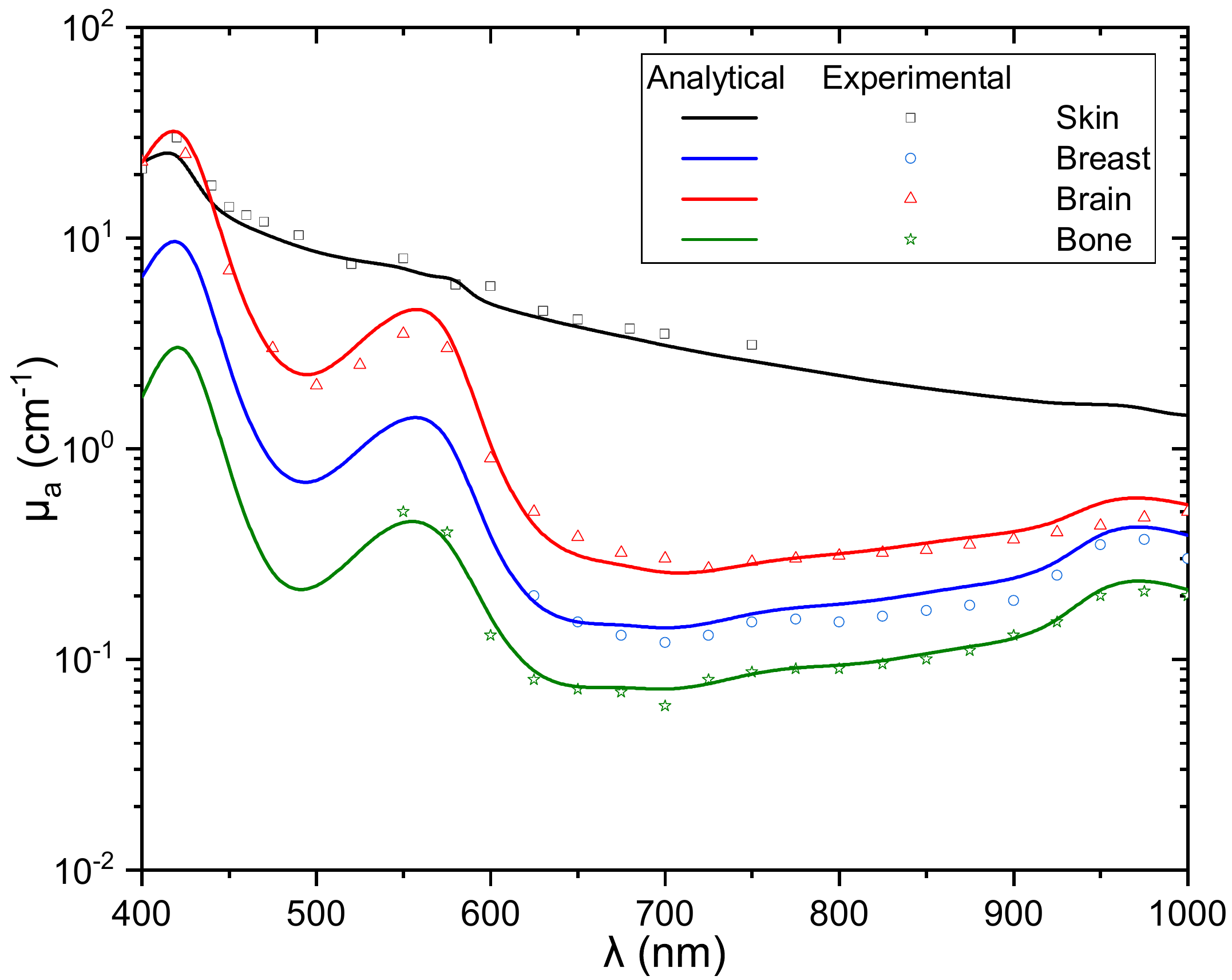}
		\caption{Absorption coefficient of complex tissues.}
		\label{fig:complex_tissues_absorption_coefficient}
	\end{subfigure}
	\begin{subfigure}[b]{0.36\textwidth}
		\vspace{0.5cm}
		\centering
		\includegraphics[width=1\linewidth,trim=0 0 0 0,clip=false]{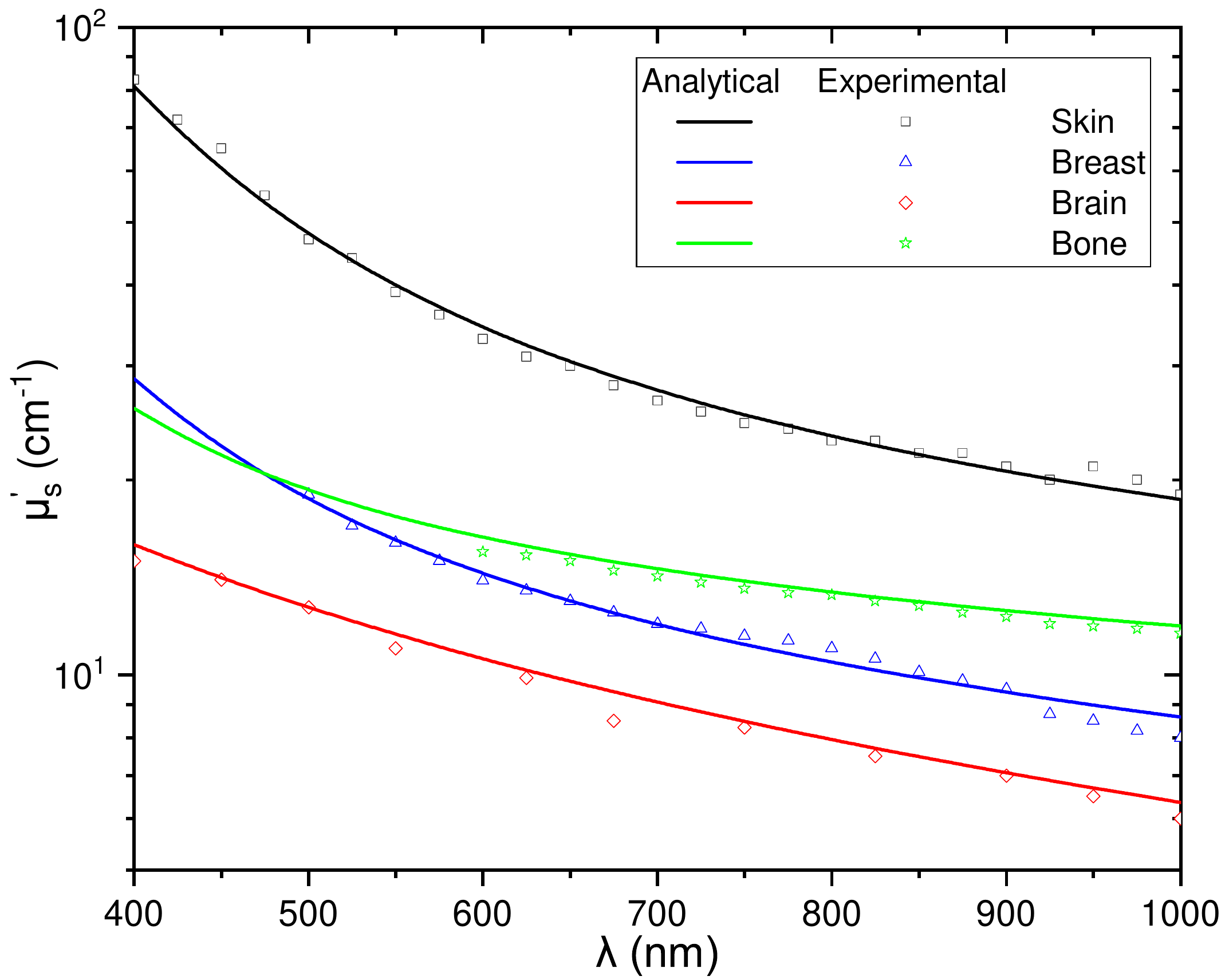}
		\caption{Reduced scattering coefficient of complex tissues.}
		\label{fig:complete_reduced_scattering_coefficient}
	\end{subfigure}
	\caption{Attenuation coefficients of complex tissues and their constituents as a function of $\lambda$.}
	\label{fig:verification}
\end{figure}

\subsection{Verification} \label{S:Attenuation}
In this subsection we verify the validity of the extracted analytical expressions presented in Section~\ref{S:pathloss_model} via comparing them with experimentally verified data from published works. In the following figures, the analytical expressions and the experimental results are depicted as continuous lines and geometric symbols, respectively. Starting with Fig.~\ref{fig:complete_fitting}, experimental data for the absorption coefficient of each constituent are plotted against the analytical results extracted from~\eqref{eq:m_oBlood} through~\eqref{eq:m_melanin}. In particular, the experimental data for the oxygenated and de-oxygenated blood, water, melanin, and fat are represented by square, circle, star, triangle, and cross symbols, respectively, while the corresponding analytical expressions are drawn in black, red, green, blue, and purple color. The validity of the proposed framework is proven based on the fact that the analytical and experimental results coincide. Moreover, the absorption coefficient of blood is not only higher than the rest constituents between $400$ and $600\text{ }\mathrm{nm}$, but also among the most influential regardless of the wavelength. Therefore, even with relatively low volume fraction, blood plays an important role in the total absorption coefficient of any generic tissue. Furthermore, it is evident that as $\lambda$ increases, the absorption coefficient increases as well, which highlights the importance of carefully selecting the transmission wavelength for tissues that are rich in water. In addition, the absorption coefficient of fat has a somewhat stable impact throughout the visible spectrum due to the fact that it receives values around $1\text{ }\mathrm{cm}^{-1}$ with very small variations. Lastly, the absorption of melanin is among the highest between the generic tissue constituents as well as the most consistent. This illustrates the significance of the concentration of melanin in the tissue under investigation and, at the same time, the negligible effect of the transmission wavelength on the absorption due to melanin. 

It should be highlighted that, the volume fraction of any of the constituents plays a very important role in the final form of the absorption coefficient. For instance, if a tissue has a high concentration in water, the impact of its absorption coefficient after it is multiplied by the water volume fraction can affect the total absorption coefficient significantly, even if the absorption coefficient of water seems insignificant on its own. On the other hand, the impact of a constituent with high absorption coefficient, such as melanin, can be diminished if it has a low volume fraction. Thus, although the absorption coefficients presented in~Fig.~\ref{fig:complete_fitting} play an important role in determining which of them affect the total absorption coefficient of the generic tissue, it is not an absolute metric. The rest of this subsection illustrates the performance of the proposed mathematical framework by comparing the estimation of the absorption and scattering coefficients of complex human tissues with experimental results from the open literature.

Next, Fig.~\ref{fig:complex_tissues_absorption_coefficient} presents the absorption coefficients of complex tissues, such as skin, bone, brain and breast, as a function of the transmission wavelength. The parameters that characterise the constitution of each tissue are provided in Table~\ref{tbl:tissue_properties} alongside their sources. From Fig.~\ref{fig:complex_tissues_absorption_coefficient}, it can be observed that the analytical expression for the total absorption coefficient provide a very close fit to the experimental data, which not only verifies the extracted expressions but also complements the accuracy of the presented mathematical framework in describing the optical absorption in generic tissues. Moreover, the absorption coefficient of the skin has the most linear behavior out of all the plotted tissues. This happens due to the increased concentration of melanin in the skin, which leads to increased absorption in higher wavelengths. In addition, the absorption coefficients of other tissues that are rich in blood and water, bare a strong resemblance to the blood absorption coefficient in the region between $400$ and $600\text{ }\mathrm{nm}$, while the impact of water becomes visible after $900\text{ }\mathrm{nm}$. The higher concentration of blood and water results in increased attenuation in the wavelengths where each of them has relatively high absorption.
\begin{table}[h]
	\centering
	\addtolength{\tabcolsep}{-2pt}
	\caption{Tissue parameters related to optical absorption and scattering for skin, bone, brain and breast tissue.}
	\begin{tabular}{|c||c|c|c|c|c||c|c|c|c||c|}
		\hline
		Tissue & $B$ & $S$ & $W$ & $F$ & $M$ & $f_{\text{Ray}}$ & $\beta$ & $\mu'_s \left(\delta_0\right)$ & $g$ & Source \\
		\hline
		Skin & $0.41$ & $99.2$ & $26.1$ & $22.5$ & $1.15$ & $0.409$ & $0.702$ & $48$ & $0.92$ & \cite{Tseng2011,Salomatina2006,Sandell2011,Shimojo2020} \\
		\hline
		Breast & $0.5$ & $52$ & $50$ & $13$ & $0$ & $0.288$ & $0.685$ & $18.7$ & $0.96$ & \cite{Pifferi2004,Sandell2011,Spinelli2004} \\
		\hline
		Bone & $0.15$ & $30$ & $30$ & $7$ & $0$ & $0.174$ & $0.447$ & $19.3$ & $0.93$ & \cite{Sandell2011,Bashkatov2006,Ugryumova2004} \\
		\hline
		Brain & $1.71$ & $58.7$ & $50$ & $20$ & $0$ & $0.32$ & $1.09$ & $12.72$ & $0.9$ & \cite{Zhao2005,Yaroslavsky2002,Zee1993} \\
		\hline
	\end{tabular}
	\label{tbl:tissue_properties}
\end{table}

Fig.~\ref{fig:complete_reduced_scattering_coefficient} depicts the experimental data for the reduced scattering coefficient of each of the complex tissues, namely skin, bone, brain and breast, against the analytical results extracted from~\eqref{eq:reduced_scattering_coefficient}. The experimental parameters for each of the tissues are available in Table~\ref{tbl:tissue_properties} alongside their sources. This figure verifies the validity and accuracy of the extracted expressions for the reduced scattering coefficient due to the proves fact that they provide a very close fit to the experimental data. Furthermore, all of the reduced scattering coefficients not only exhibit a more linear behavior than the corresponding absorption coefficients, but also hold significantly higher values. This highlights the detrimental effect that scattering plays in the propagation of optical radiation through the human body. Moreover, it is evident that as $\lambda$ increases, the reduced scattering coefficient decreases, which suggests that as we increase the transmission wavelength the impact of scattering could prove to diminish. For example, for skin tissue, as $\delta$ rises from $400\text{ }\mathrm{nm}$ to $1000\text{ }\mathrm{nm}$, the reduced scattering coefficient decreases by $87.5\text{ }\%$. Finally, we observe that skin causes almost half an order of magnitude higher attenuation than the rest of the tissues under investigation, while brain tissue exhibits the lowest.

\subsection{Pathloss} \label{S:Pathloss}
\begin{figure}[h]
	\centering
	\begin{subfigure}[b]{0.36\textwidth}
		\centering
		\includegraphics[width=1\linewidth,trim=0 0 0 0,clip=false]{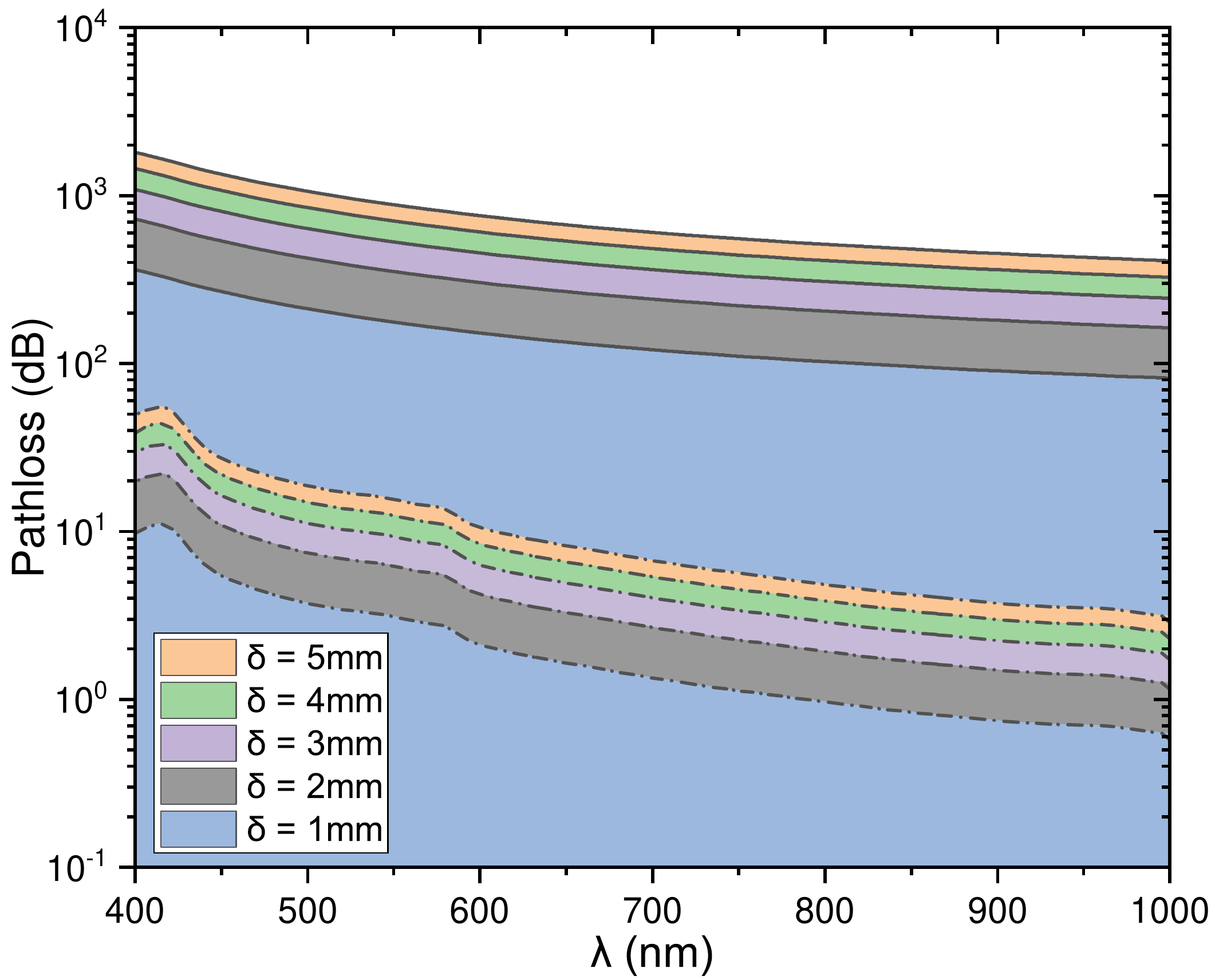}
		\caption{Skin tissue.}
		\label{fig:skin_pathloss}
	\end{subfigure}
	\begin{subfigure}[b]{0.36\textwidth}
		\vspace{0.5cm}
		\centering
		\includegraphics[width=1\linewidth,trim=0 0 0 0,clip=false]{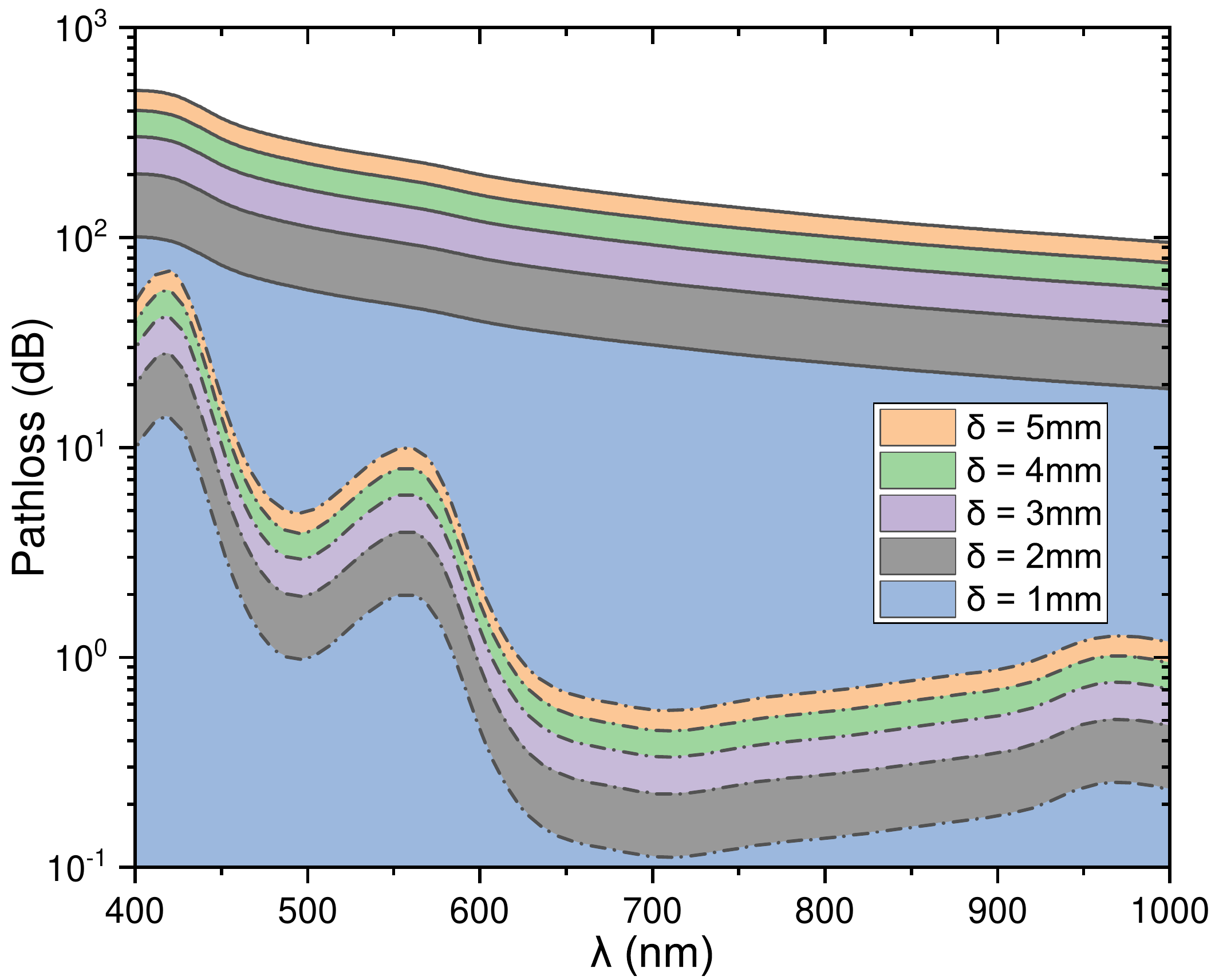}
		\caption{Brain tissue.}
		\label{fig:brain_pathloss}
	\end{subfigure}
	\begin{subfigure}[b]{0.36\textwidth}
		\centering
		\includegraphics[width=1\linewidth,trim=0 0 0 0,clip=false]{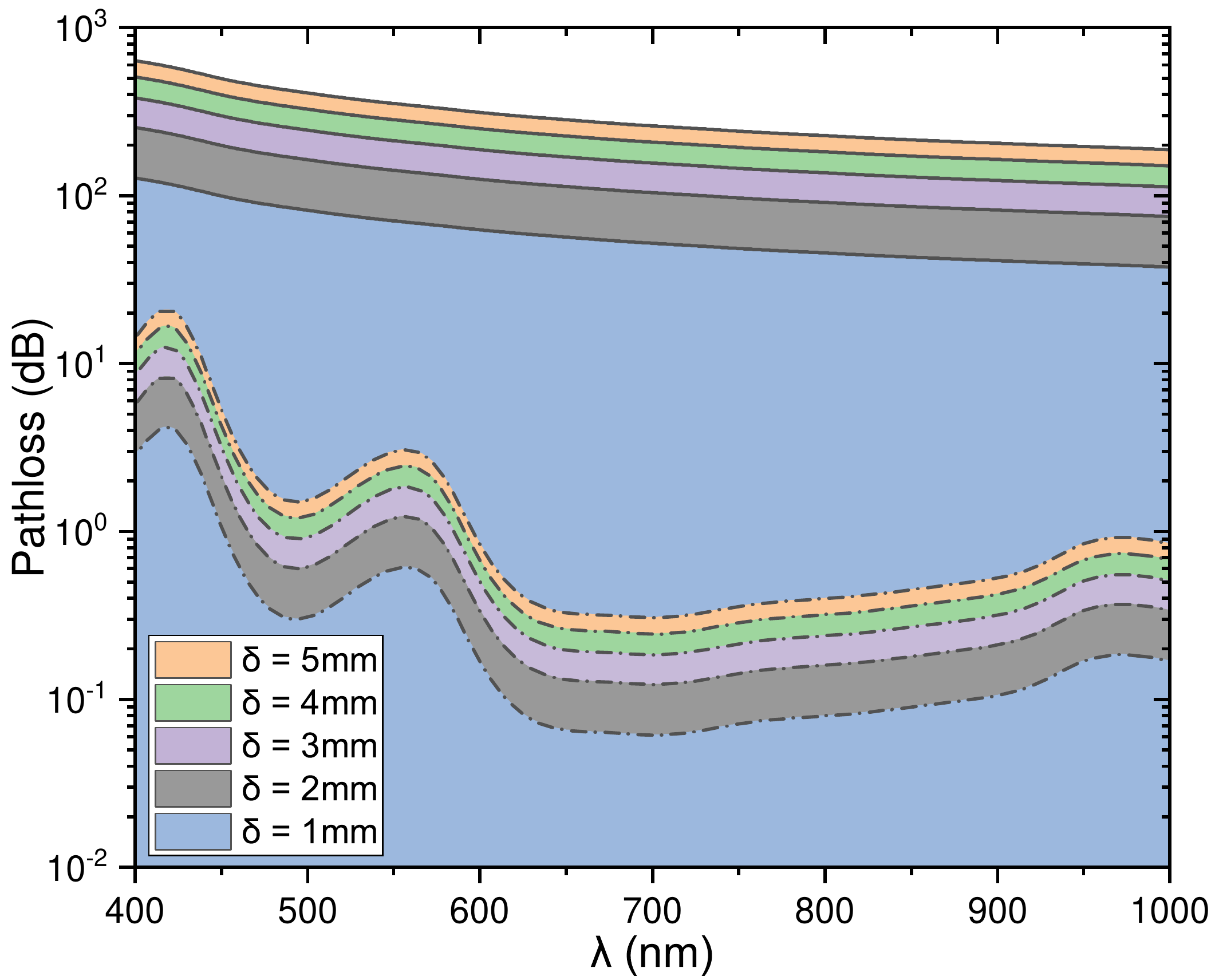}
		\caption{Breast tissue.}
		\label{fig:breast_pathloss}
	\end{subfigure}
	\begin{subfigure}[b]{0.36\textwidth}
		\vspace{0.5cm}
		\centering
		\includegraphics[width=1\linewidth,trim=0 0 0 0,clip=false]{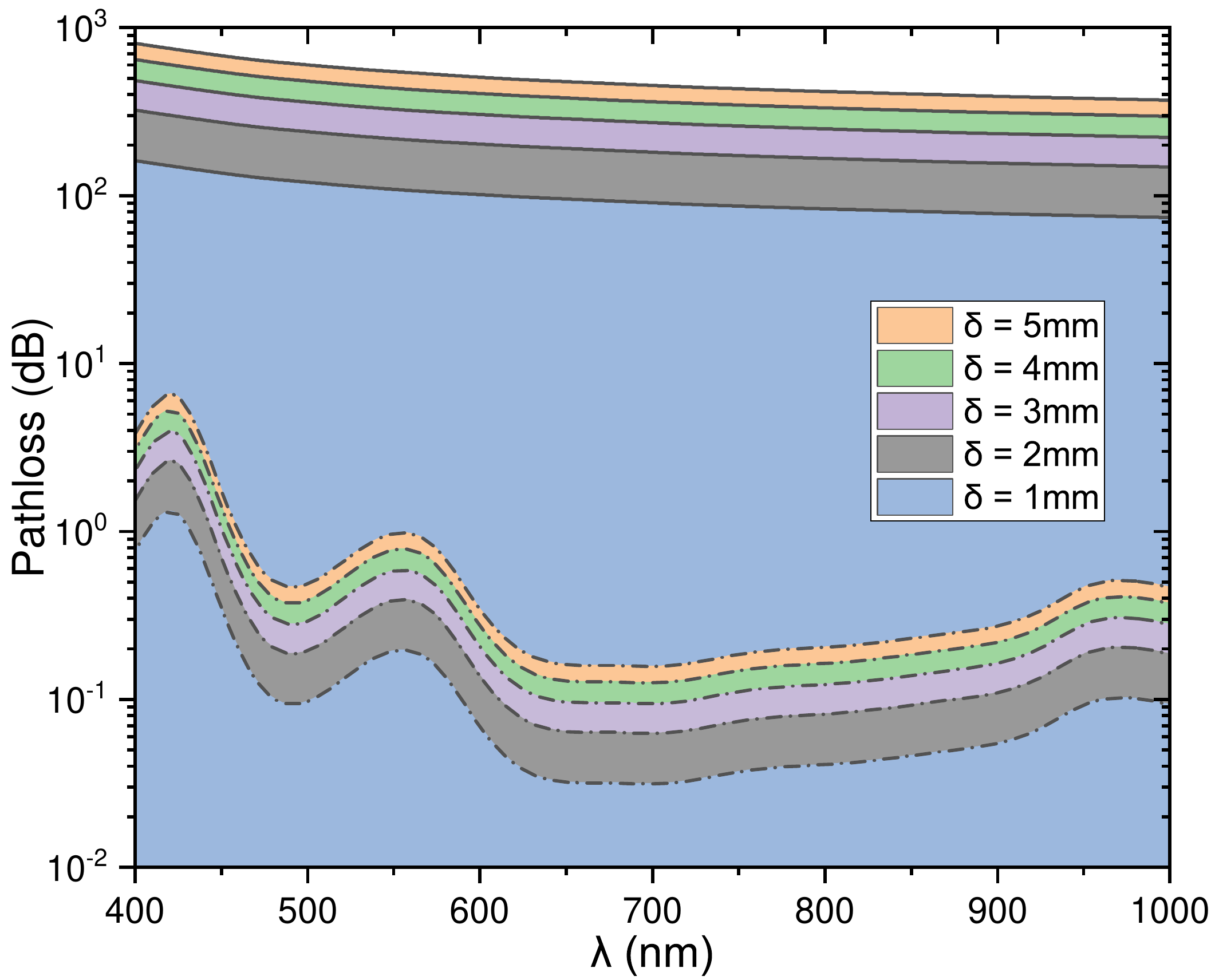}
		\caption{Bone tissue.}
		\label{fig:bone_pathloss}
	\end{subfigure}
	\caption{Pathloss due to absorption and scattering of various tissues as a function of the transmission wavelength for different values of tissue thickness. Continuous and dashed lines denote the complete pathloss and the pathloss only due to the absorption coefficient.}
\end{figure}

Based on the extracted accurate expressions for the reduced scattering and absorption coefficients of various complex tissues, we can estimate the pathloss. Therefore, Fig.~\ref{fig:skin_pathloss} depicts the pathloss as a function of the wavelength due to absorption and scattering in skin tissues with different values of thickness. From this figure, it is evident that the complete pathloss (continuous line) decreases with the wavelength and increases with the skin thickness, while the pathloss due to absorption (dashed line) has more fluctuations throughout the visible spectrum. In more detail, we observe that the optimal wavelength is $1000\text{ }\mathrm{nm}$. Furthermore, in the following we assume a transmission window to be the spectrum region where the pathloss does not exceed $6\text{ }\mathrm{dB}$, i.e. the residual optical signal is at least a quarter of the transmitted one. Thus, for $\delta = 1\text{ }\mathrm{mm}$ a transmission window exists for wavelength values higher than $450\text{ }\mathrm{nm}$, which shrinks as the transmission distance increases. For instance, for $\delta = 3\text{ }\mathrm{mm}$ it reduces to wavelengths higher than $650\text{ }\mathrm{nm}$, while for $\delta = 3\text{ }\mathrm{mm}$ it becomes even smaller for wavelengths higher than $650\text{ }\mathrm{nm}$. On the other hand, if we take into consideration the scattering coefficient, the complete pathloss exhibits a more linear behavior. Specifically, the impact of scattering in the propagation of light through biological tissues is approximately two orders of magnitude higher than that of the absorption, which constitutes if as the dominant phenomenon concerning the optical transmission. For example, if we assume $\delta = 5\text{ }\mathrm{mm}$ and $\lambda = 600\text{ }\mathrm{nm}$, the scattering coefficient exceeds $\mu_a$ $100$ times, while for $\lambda = 600\text{ }\mathrm{nm}$, $\mu_s$ is $125$ times higher. As a result, it is obvious that the absorption coefficient decreases with an increased rate compared to $\mu_s$.

Fig.~\ref{fig:brain_pathloss} illustrates the pathloss due to the existence of brain tissue as a function of the wavelength for different values of $\delta$. As expected, for higher values of $\delta$ the pathloss is also higher, while the behavior of pathloss for wavelength variations is not linear. For example, as $\lambda$ increases from $500$ to $550\text{ }\mathrm{nm}$ the pathloss increases, while for the same increase from $550$ to $600\text{ }\mathrm{nm}$, pathloss decreases. Also, the optimal transmission wavelength is $700\text{ }\mathrm{nm}$. In addition, two transmission windows exist for $\delta = 1\text{ }\mathrm{mm}$. The first is between $450$ and $550\text{ }\mathrm{nm}$, while the second after $600\text{ }\mathrm{nm}$. However, as the transmission distance increases, only the second window will be valid. Furthermore, if we take into consideration the effect of scattering, the complete pathloss takes much higher values and the fluctuations of the absorption become almost obsolete, with its impact being visible only in the $400$ to $500\text{ }\mathrm{nm}$ range. By observing Fig.~\ref{fig:brain_pathloss}, it becomes evident that the impact of $\mu_a$ varies greatly with the wavelength, which results in fluctuations of the relation between $\mu_s$ and $\mu_a$. For instance, for $\delta = 1\text{ }\mathrm{mm}$, as $\lambda$ increases from $400$ to $700\text{ }\mathrm{nm}$, the relation between the scattering and absorption coefficient increases from $10x$ to $400x$.

In Fig.~\ref{fig:breast_pathloss}, the pathloss for transmission through breast tissue is presented with regard to the transmission wavelength for various values of tissue thickness. If we take into account only the absorption phenomenon, it is evident that, as the $\delta$ increases, the pathloss increases as well. On the contrary, the wavelength influences the pathloss in a non linear manner. Also, a single transmission window os observed for all the plotted values of $\delta$ for wavelengths higher than $550\text{ }\mathrm{nm}$. Although the optimal transmission wavelength for breast tissue with regard to the absorption is $700\text{ }\mathrm{nm}$, after introducing the scattering coefficient into the analysis, the complete pathloss exhibits an almost linear behavior and the optimal wavelength is shifted to $1000\text{ }\mathrm{nm}$. The increased impact of $\mu_s$ varies between $1$ and $3$ orders of magnitude more than the absorption coefficient's. Specifically, for $\delta = 1\text{ }\mathrm{mm}$ and $\lambda = 400\text{ }\mathrm{nm}$, $\mu_s$ is approximately $33$ times higher than $\mu_a$, while for $\lambda = 650\text{ }\mathrm{nm}$, it is $600$ times higher.

Finally, Fig.~\ref{fig:bone_pathloss} depicts the pathloss through bone tissue as a function of the transmission wavelength for various transmission distance values. Yet again, pathloss increases with tissue thickness, while its behavior with regard to $\lambda$ depends. The optimal $\lambda$ for transmission through bone tissue is $700\text{ }\mathrm{nm}$. For $\delta$ equal to $1$ and $2\text{ }\mathrm{mm}$, only one transmission window exists for $\lambda$ higher than $450\text{ }\mathrm{nm}$. On the contrary, for higher $\delta$ values, there are two transmission windows. For example, for $\delta = 5\text{ }\mathrm{mm}$, the two windows are $475 - 525\text{ }\mathrm{nm}$, and $600 - 950\text{ }\mathrm{nm}$. However, the complete pathloss appears to be the most detrimental between the studied tissues, which is expected due to the more compact constitution of the bone tissue. This effect is highlighted by the fact that after adding the scattering phenomenon in our simulations the complete pathloss is increased by almost three orders of magnitude. In more detail, for a specific value of $\delta$ equal to $5\text{ }\mathrm{mm}$, as the transmission wavelength increases from $400$ to $950\text{ }\mathrm{nm}$, the relation between the $\mu_s$ and $\mu_a$ increases from $200x$ to $2000x$.

\section{Conclusions} \label{S:conclusion}
This paper introduces a novel mathematical framework that models the optical signal's attenuation as it travels through any generic biological tissue. Initially, we extract analytical expressions for the absorption coefficient of the major generic tissue constituents based on published experimental measurements, which enable the estimation of not only the absorption coefficient of each constituent at any given wavelength but also the absorption coefficient of any generic tissue. Moreover, the phenomenon of scattering is incorporated into the proposed framework as a complex stochastic process comprised of a Rayleigh and a Mie scattering component in order to model the impact due to the existence of smaller and larger particles, respectively, in the generic tissue. Next, the phenomena of absorption and scattering are incorporated into a unified framework, which is validated by comparing the analytical results with experimental data from the open literature. Finally, we illustrate the pathloss as a function of the transmission wavelength for different complex tissues and tissue thickness, and provide insightful discussions as well as design guidelines for future in-body OWC applications.

\vspace{6pt} 



\authorcontributions{Conceptualization, S.E.T. and Y.Y.; methodology, S.E.T; software, S.E.T; validation, S.E.T, A.-A.A.B. and N.D.C.; writing---original draft preparation, S.E.T.; writing---review and editing, A.-A.A.B. and N.D.C.; visualization, S.E.T.; supervision, A.-A.A.B.; project administration, N.D.C.; funding acquisition, N.D.C. All authors have read and agreed to the published version of the manuscript.}

\funding{This research is co-financed by Greece and the European Union (European Social Fund-ESF) through the Operational Programme “Human Resources Development, Education and Lifelong Learning 2014-2020” in the context of the project “IRIDA-Optical wireless communications for in-body and transdermal biomedical applications” (MIS 5047929).}

\conflictsofinterest{The authors declare no conflict of interest.} 


%

\end{paracol}
\reftitle{References}


\externalbibliography{yes}
\bibliography{mybibfile}

\section*{Short Biography of Authors}
\bio
{\raisebox{-0.35cm}{\includegraphics[width=3.5cm,height=5.3cm,clip,keepaspectratio]{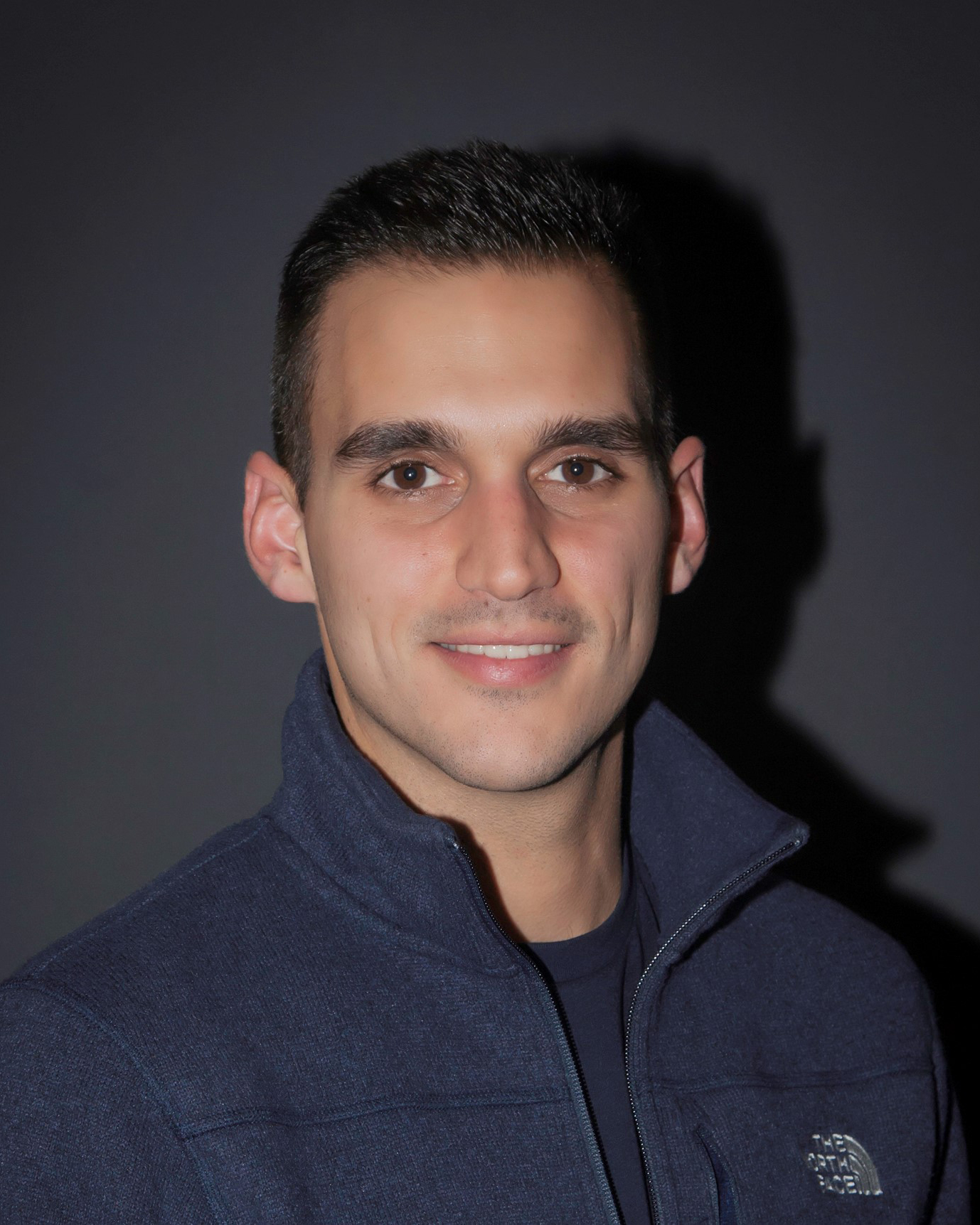}}}
{\textbf{Stylianos E. Trevlakis} was born in Thessaloniki, Greece in 1991. He received the Electrical and Computer Engineering (ECE) diploma (5 year) from the Aristotle University of Thessaloniki (AUTh) in 2016. Afterwards he served in the Hellenic Army in for nine months in the Research Office as well as at the Office of Research and Informatics of the School of Management and Officers. During 2017, he joined the Information Technologies Institute, while from October 2017, he joined WCSG underthe leadership Prof. Karagiannidis in AUTh as a PhD candidate. \\ \\
His research interests are in the area of Wireless Communications, with emphasis on Optical Wireless Communications, and Communications \& Signal Processing for Biomedical Engineering.}

\bio
{\raisebox{-0.35cm}{\includegraphics[width=3.5cm,height=5.3cm,clip,keepaspectratio]{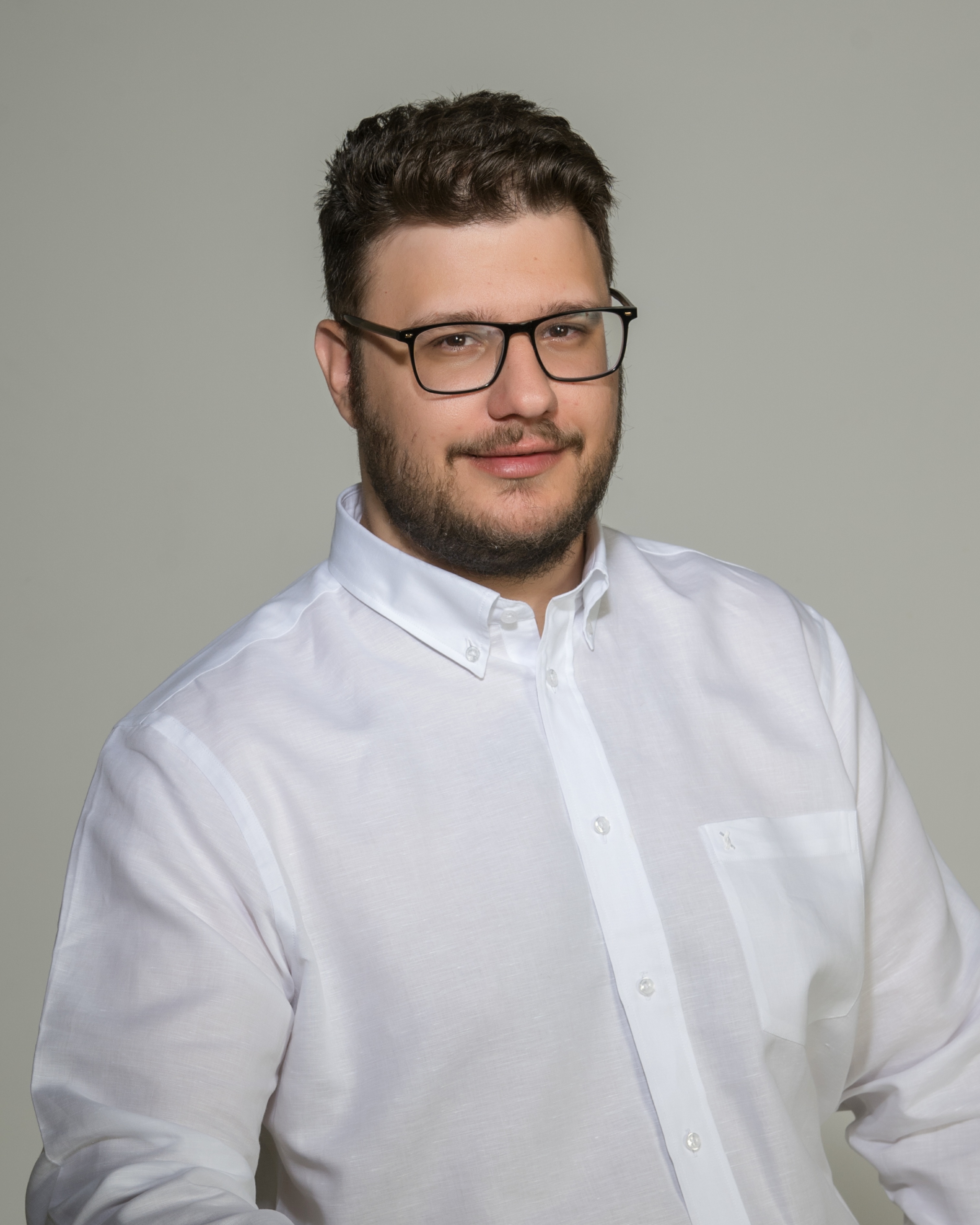}}}
{\textbf{Alexandros-Apostolos A. Boulogeorgos} (S’11, M’16, SM’19) was born in Trikala, Greece in 1988. He received the Electrical and Computer Engineering (ECE) diploma degree and Ph.D. degree in Wireless Communications from the Aristotle University of Thessaloniki (AUTh) in 2012 and 2016, respectively. From November 2012, he joined the wireless communications system group of AUTh, where he worked as a research assistant/project engineer in various telecommunication and networks projects. During 2017, he joined the information technologies institute (ITI) of the Center for Research \& Technology Hellas (CERTH), where he conducted research in the topics of wireless technologies for internet-of-thing applications. From November 2017, he has joined the Department of Digital Systems, University of Piraeus. From October 2012 until September 2016, he was a teaching assistant at the department of ECE of AUTh, whereas, from February 2017, he serves as an adjunct professor at the Department of ECE of the University of Western Macedonia, Greece. Likewise, he serves as a visiting lecturer in post-graduate courses at the University of Thessaly, Greece. \\ \\
Dr. Boulogeorgos has authored and co-authored more than 75 technical papers, which were published in scientific journals and presented at prestigious international conferences. Furthermore, he has submitted two (one national and one European) patents. Additionally, he has been involved as member of Organization and Technical Program Committees in several IEEE and non-IEEE conferences and served as a reviewer in various IEEE and non-IEEE journals and conferences. He was awarded with the “Distinction Scholarship Award” of the Research Committee of AUTh for the year 2014 and was recognized as an exemplary reviewer for IEEE Communication Letters for 2016 (top 3\% of reviewers). Moreover, he was named a top peer reviewer (top 1\% of reviewers) in Cross-Field and Computer Science in the Global Peer Review Awards 2019, which was presented by the Web of Science and Publons. In 2021, he received the best presentation award in the International Conference on Modern Circuits and Systems Technologies (MOCAST). \\ \\
His current research interests spans in the area of wireless communications and networks with emphasis in high frequency communications, optical wireless communications and communications for biomedical applications. He is a Senior Member of the IEEE and a member of the Technical Chamber of Greece. He is currently an Editor for IEEE Communications Letters, an Associate Editor for the Frontier In Communications And Networks, and an Editor in MDPI Telecoms.}

\bio
{\raisebox{-0.35cm}{\includegraphics[width=3.5cm,height=5.3cm,clip,keepaspectratio]{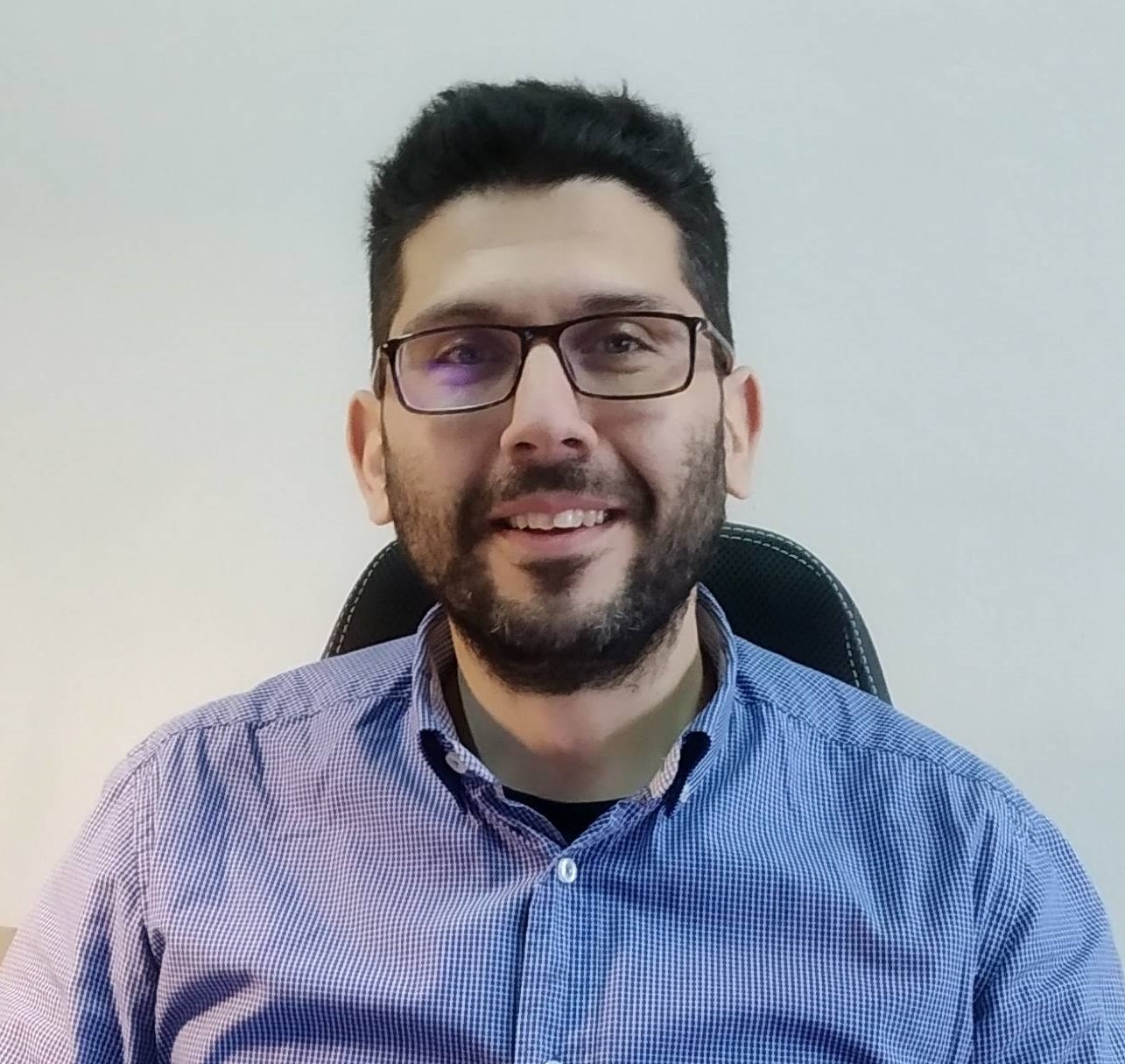}}}
{\textbf{Nestor D. Chatzidiamantis} (S’08, M’14) was born in Los Angeles, CA, USA, in 1981. He received the Diploma degree (5 years) in electrical and computer engineering (ECE) from the Aristotle University of Thessaloniki (AUTH), Greece, in 2005, the M.Sc. degree in telecommunication networks and software from the University of Surrey, U.K., in 2006, and the Ph.D. degree from the ECE Department, AUTH, in 2012. From 2012 through 2015, he worked as a Post-Doctoral Research Associate in AUTH and from 2016 to 2018, as a Senior Engineer at the Hellenic Electricity Distribution Network Operator (HEDNO). Since 2018, he has been Assistant Professor at the ECE Department of AUTH and member of the Telecommunications Laboratory.\\ \\
His research areas span signal processing techniques for communication systems, performance analysis of wireless communication systems over fading channels, communications theory, cognitive radio and free-space optical communications.}

\bio
{\raisebox{-0.35cm}{\includegraphics[width=3.5cm,height=5.3cm,clip,keepaspectratio]{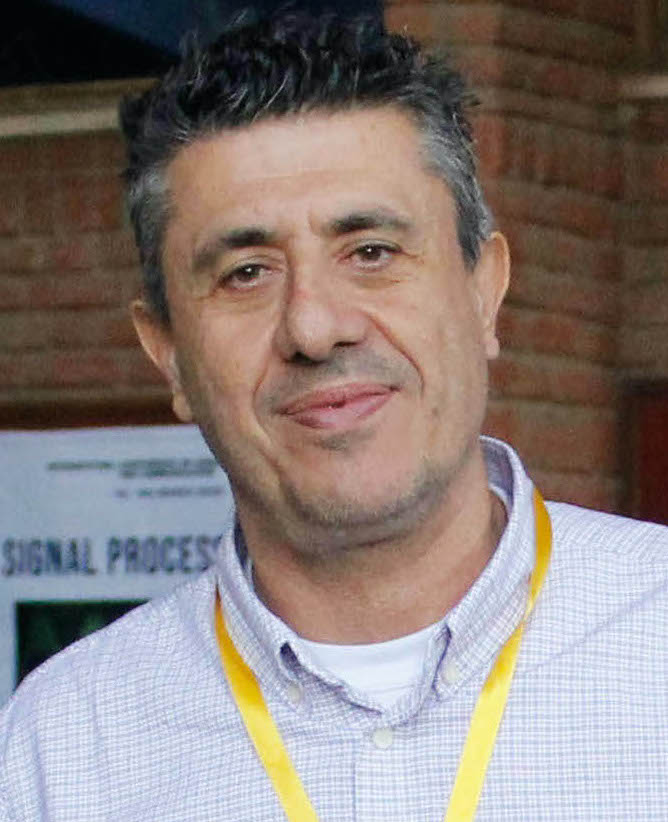}}}
{\textbf{George K. Karagiannidis} (M'96-SM'03-F'14) was born in Pithagorion, Samos Island, Greece. He received the University Diploma (5 years) and PhD degree, both in electrical and computer engineering from the University of Patras, in 1987 and 1999, respectively. From 2000 to 2004, he was a Senior Researcher at the Institute for Space Applications and Remote Sensing, National Observatory of Athens, Greece. In June 2004, he joined the faculty of Aristotle University of Thessaloniki, Greece where he is currently Professor in the Electrical \& Computer Engineering Dept. and Head of Wireless Communications \& Information Processing Systems Group (WCIP).  He is also Honorary Professor at South West Jiaotong University, Chengdu, China.\\ \\ 	
His research interests are in the broad area of Digital Communications Systems and Signal processing, with emphasis on Wireless Communications, Optical Wireless Communications, Wireless Power Transfer and Applications and Communications \& Signal Processing for Biomedical Engineering.\\ \\ 	
Dr. Karagiannidis has been involved as General Chair, Technical Program Chair and member of Technical Program Committees in several IEEE and non-IEEE conferences. In the past, he was Editor in several IEEE journals and from 2012 to 2015 he was the Editor-in Chief of IEEE Communications Letters. Currently, he serves as Associate Editor-in Chief of IEEE Open Journal of Communications Society.\\ \\	
Dr. Karagiannidis is one of the highly-cited authors across all areas of Electrical Engineering, recognized from Clarivate Analytics as Web-of-Science Highly-Cited Researcher in the five consecutive years 2015-2019.}

\end{document}